\documentclass[12pt]{iopart}
\usepackage{graphicx}
\usepackage[usenames]{color}
\usepackage[dvips]{epsfig}
\usepackage{bbm}
\usepackage{verbatim}
\usepackage{psfrag}
\usepackage{slashed}

\def \Ns {N_{\sigma}}
\def \Nt {N_{\tau}}

\newcommand{\Exp}[1]{\exp \lr{#1}}

\newcommand{\lr}[1]{\left( #1 \right)}

\newcommand{\beqn} {\begin{equation}}
\newcommand{\eqn} {\end{equation}}

\def \beq{\begin{equation}}
\def \eeq{\end{equation}}
\def \bea{\begin{eqnarray}}
\def \eea{\end{eqnarray}}

\def \bet0{\beta_0}
\def \bet1{\beta_1}
\def \simgt{\,\rlap{\lower 7.5 pt\hbox{$\mathchar \sim$}}\raise 3 pt \hbox{$>$}\,}
\def \simlt{\,\rlap{\lower 7.5 pt\hbox{$\mathchar \sim$}}\raise 3 pt \hbox{$<$}\,}

\def\lsim{\raise0.3ex\hbox{$<$\kern-0.75em\raise-1.1ex\hbox{$\sim$}}}
\def\gsim{\raise0.3ex\hbox{$>$\kern-0.75em\raise-1.1ex\hbox{$\sim$}}}

\begin{document}

\vspace*{-2.5cm}
\begin{flushright}
\texttt{\footnotesize CERN-PH-TH/2011-158}\\
\end{flushright}

\title{Continuous Time Monte Carlo for Lattice QCD \break
in the Strong Coupling Limit}
\author{
W.~Unger$^{\rm a}$,
Ph.~de Forcrand$^{\rm a,b}$
}
\address{
$^{\rm a}$ Institut f\"ur Theoretische Physik, ETH Z\"urich, CH-8093 Z\"urich, Switzerland\\
$^{\rm b}$ CERN, Physics Department, TH Unit, CH-1211 Geneva 23, Switzerland \\
}

\begin{abstract}
We present results for lattice QCD in the limit of infinite gauge coupling, obtained from a worm-type Monte Carlo algorithm on a discrete spatial lattice but with continuous Euclidean time. This is obtained by sending both the anisotropy parameter $\gamma^2\simeq a/a_t$ and the number of time-slices $N_\tau$ to infinity, keeping the ratio $\gamma^2/N_\tau \simeq aT$ fixed. The obvious gain is that no continuum extrapolation $N_\tau \rightarrow \infty$ has to be carried out. Moreover, the algorithm is faster and the sign problem disappears. We compare our computations with those on discrete lattices. We determine the phase diagram as a function of temperature and baryon chemical potential. 
\end{abstract}


\section{Introduction}

The determination of the QCD phase diagram, in particular the location of the critical point, is an important, long standing problem, requiring non-perturbative methods.
In lattice QCD, several approaches have been developed to investigate the phase transition from the hadronic matter to the quark gluon plasma,
but all of them are limited to $\mu_B/T \lsim 1$, with $\mu_B$ the baryon chemical potential.
The reason for this is the notorious sign problem, which arises because the fermion determinant for finite $\mu$ becomes complex, and importance sampling is no longer applicable. 
In QCD, the sign problem is severe, since the relative fluctuations of the complex phase factor grow exponentially with the lattice volume.
In the strong coupling limit of lattice QCD (SC-QCD) discussed below, the sign problem is however mild and the full $(\mu,T)$ phase diagram can be obtained.

\section{Strong Coupling Lattice QCD}

In SC-QCD, the gauge coupling is sent to infinity and hence the coefficient of the plaquette term $\beta=6/g^2$ is sent to zero. Hence the Yang Mills part $F_{\mu\nu} F_{\mu\nu}$ is absent. Then, the gauge fields in the covariant derivative can be integrated out analytically. 
However, as a consequence of the strong coupling limit, the lattice spacing becomes very large, and no continuum limit can be performed.
The degrees of freedom in SC-QCD live on a crystal.
We consider the SC-limit for staggered fermions
as studied in \cite{Karsch1989}, which after gauge link integration and Grassman integration of staggered fermion fields yields a partition function that describes a system of confined, colorless, discrete degrees of freedom:

\begin{itemize} 
\item Mesonic degrees of freedom $k_{\hat{\mu}}(x)\in \{0,\ldots 3\}$ (non-oriented meson hoppings called dimers) and
$n(x) \in \{0,\ldots 3\}$ (mesonic sites called monomers), which obey the Grassmann constraint $n(x)+\sum_{\hat{\mu}=\pm\hat{0},\ldots \pm \hat{d}} k_{\hat{\mu}}(x)=3$ at every lattice site $x\in \Ns^3\times \Nt$;
\item Baryonic degrees of freedom, which form oriented baryon loops. These loops are self-avoiding and do not touch the mesonic degrees of freedom.
\end{itemize}
The weights for temporal meson or baryon hoppings contain the anisotropy parameter $\gamma$, needed in order to vary the temperature $aT\simeq\gamma^2/\Nt$ continuously (see below).
Here, we will restrict to the chiral limit, $m_q=0$. In that case monomers are absent.

\section{Continuum Limit and Continuous Time Worm Algorithm}
\begin{figure}[b!]
\vspace{-5mm}
\begin{minipage}{0.69\textwidth}
\includegraphics[width=\textwidth]{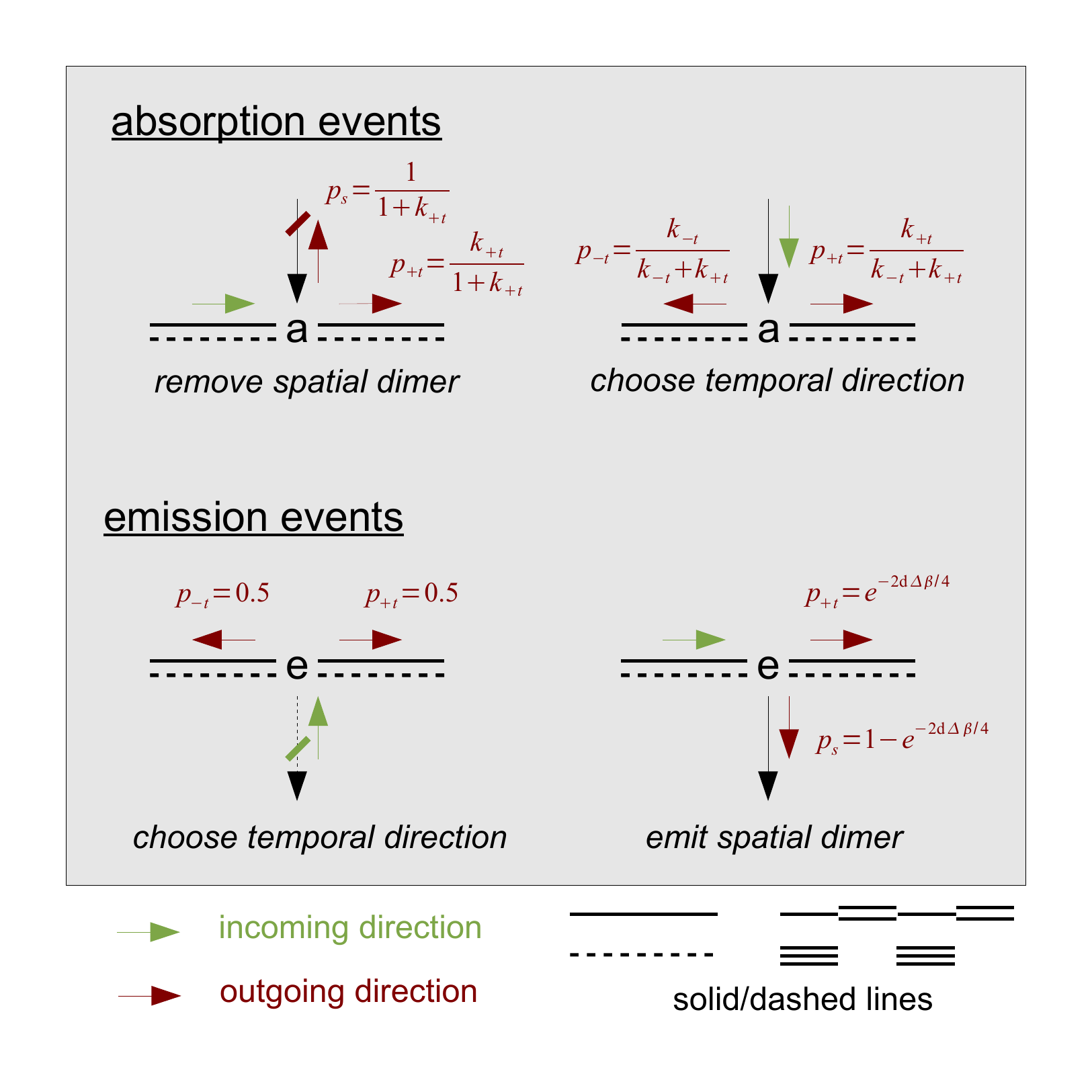}
\end{minipage}
\begin{minipage}{0.3\textwidth}
\includegraphics[width=\textwidth]{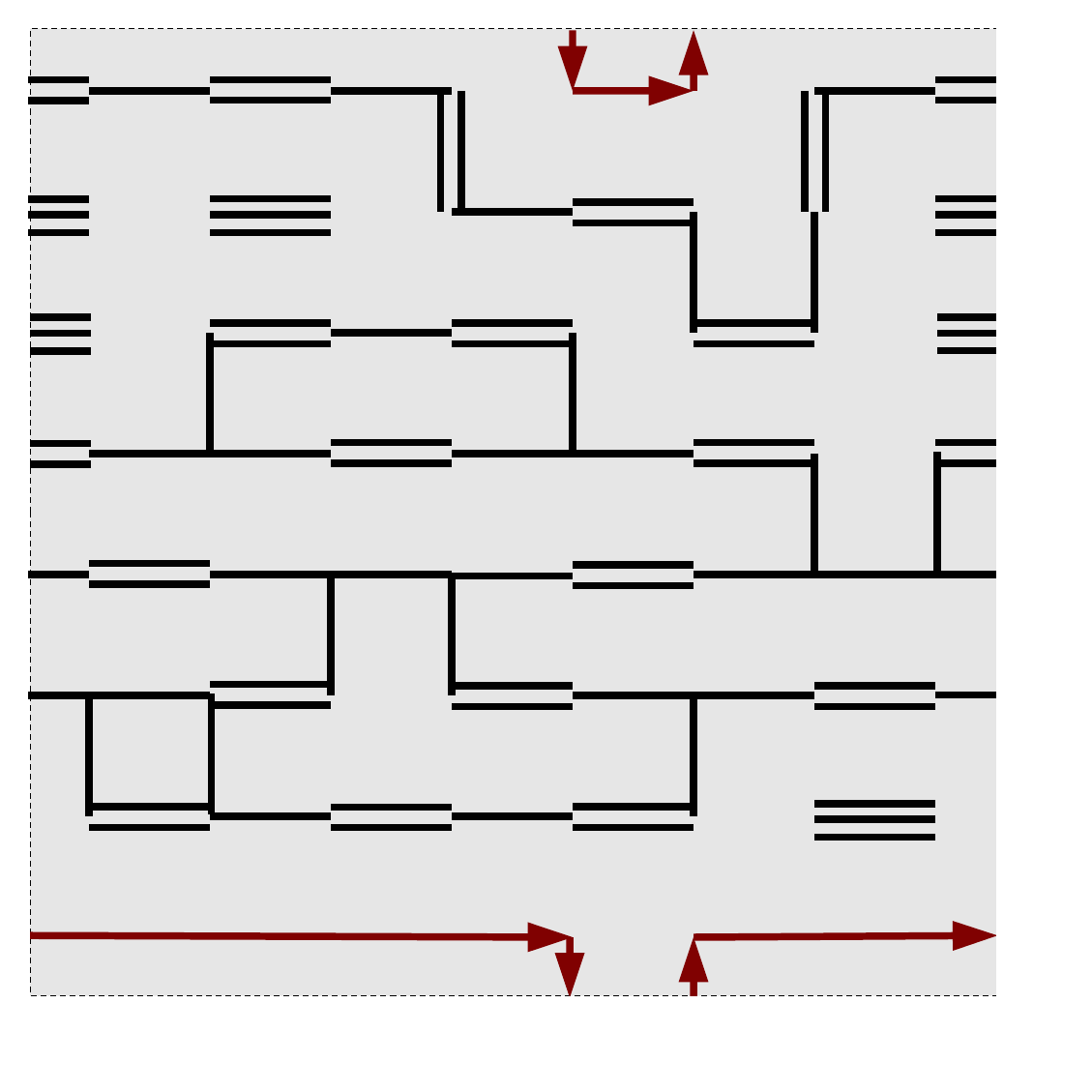}
\includegraphics[width=\textwidth]{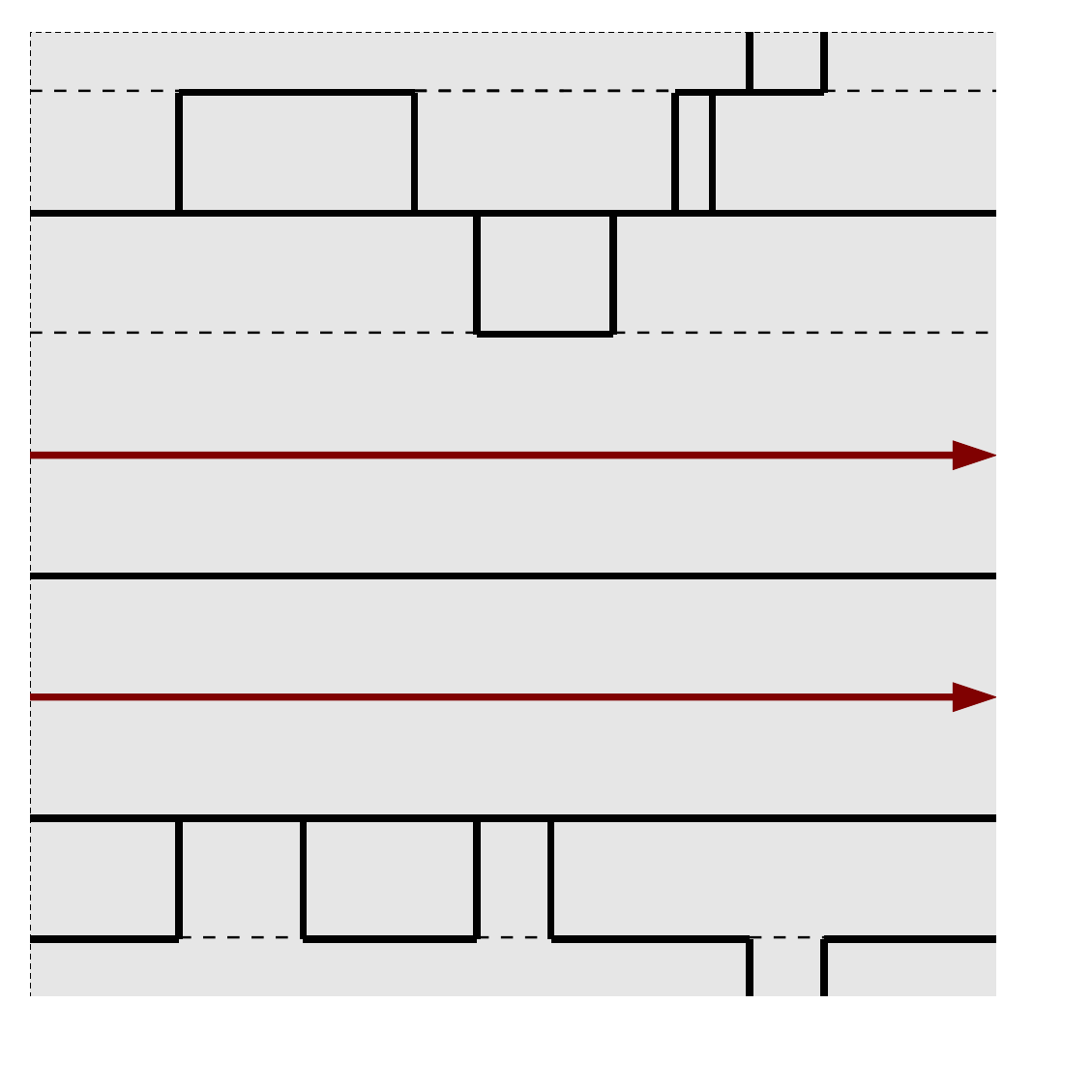}
\end{minipage}
\vspace{-5mm}
\caption{\emph{Left:} updating rules for the continuous time algorithm. \emph{Right:} illustrative 2-dim.~configurations (time flows to the right), in discrete time ({\em top}) and continuous time ({\em bottom}). Note how the latter lacks multiple spatial dimers and has only static baryons lines.
\vspace{-15mm}\\
}
\label{absorptionemission}
\end{figure}

\begin{figure}
\includegraphics[width=0.49\textwidth]{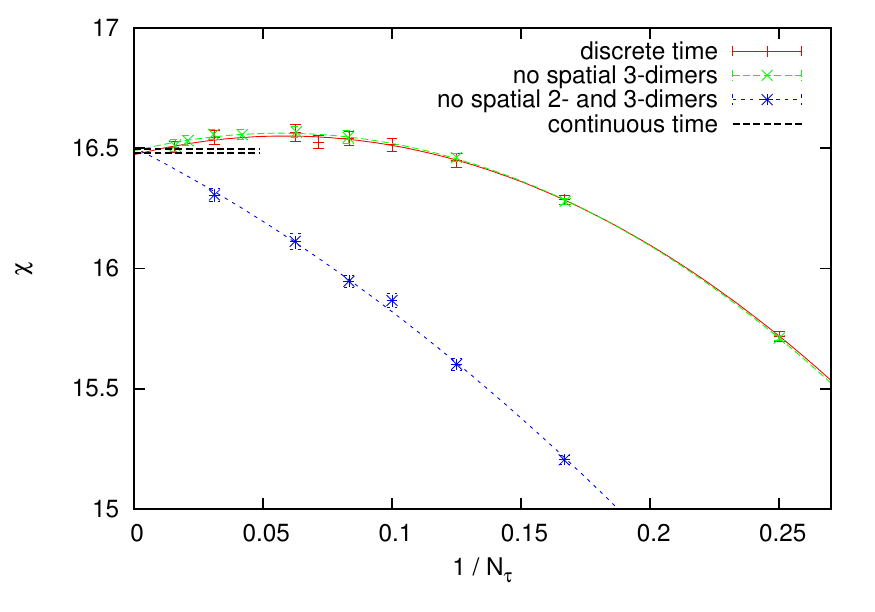}
\includegraphics[width=0.49\textwidth]{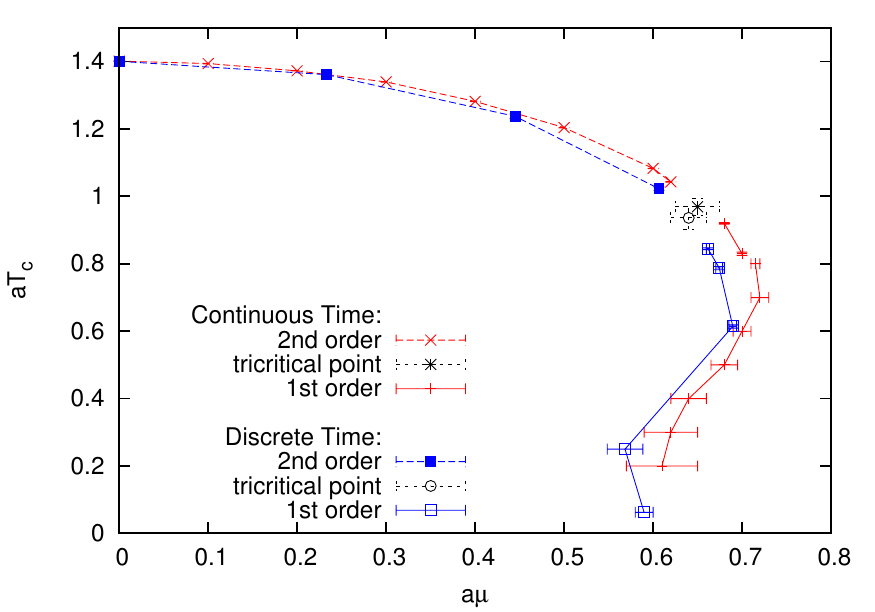}
\caption{\emph{Left:} continuous-time limit of the chiral susceptibility $\chi$ at $aT=1.8$ in the purely mesonic system U(3), exhibiting typical non-monotonic behavior in $1/\Nt$. Note that the spatial suppression of double and/or triple spatial dimers has no effect on the CT-limit.
\emph{Right:} The SC-QCD phase diagram obtained on an $\Nt =4$ lattice~\cite{Forcrand2010} and in continuous time.}
\label{U3}
\vspace{5mm}
\end{figure}

Continuous time (CT) algorithms are now widely used in quantum Monte Carlo (see e.g. \cite{Beard1996, Gull2010}), but to our knowledge have not yet been applied to quantum field theories.
The continuum limit in Euclidean time which we are interested in is the limit 
\vspace{2mm}
\begin{equation}
\Nt\rightarrow \infty, \qquad \gamma \rightarrow \infty, \qquad \gamma^2/\Nt\quad {\rm fixed}
\end{equation}
\vspace{2mm}
as $ \gamma^2/\Nt$ represents the temperature $aT$.
Designing an algorithm that operates in this limit will have several advantages: 
There is no need to perform the continuum extrapolation $\Nt \rightarrow \infty$, which 
allows to estimate critical temperatures more precisely, with a faster algorithm. Moreover, ambiguities arising from the functional dependence of observables on the anisotropy parameter (esp. non-monotonic behaviour as in Fig.~\ref{U3} left) will be circumvented.
Also in the baryonic part of the partition function great simplifications occur: Baryons become static in the CT-limit, hence the sign problem is completely absent.
The partition function can be written in terms of vertices at which spatial meson hoppings occur:
\vspace{2mm}
\begin{equation}
\mathcal{Z}(\gamma, \Nt) 
\simeq 
\sum_{\{k,\sigma\}} \prod_{x\in V_M} v_L^{n_L(x)} v_T^{n_T(x)}\prod_{x\in V_B} \Exp{-3\sigma(x) \mu_q/T}
\label{PARF}
\end{equation}
\vspace{2mm}
This relation becomes exact in the CT-limit since spatial dimers with multiplicity 2 or 3 are suppressed by powers of $\gamma$ and are hence absent at $\gamma\rightarrow \infty$, see Fig.~\ref{absorptionemission} right: as the temporal lattice spacing $a_t\simeq a/\gamma^2 \rightarrow 0$, multiple spatial dimers become resolved into single dimers. 
The overall number of spatial dimers remains finite in the CT-limit, as the sum over ${\cal O}(\gamma^2)$ sites compensates the $1/\gamma^2$ suppression. 
Temporal dimers can be arranged in chains of alternating 3-dimers and 0-dimers, which we denote by dashed lines, and 2-dimers and 1-dimers, which we denote by solid lines (see Fig.~\ref{absorptionemission}). The crucial observation is that the weight of these chains in the partition function is independent of their length, as the weight of each 3-dimer cancels that of the 0-dimer and likewise the weight of 2-dimers cancels that of 1-dimers. Hence, the weight of a configuration will only depend on the kind and number of vertices at which spatial hoppings are attached to solid/dashed lines, not on their position. For SC-QCD, there are two kinds of vertices, ``L''-vertices of weight $v_L=\gamma^{-1}$, where dashed and solid lines join, and ``T''-vertices of weight $v_T=2\gamma^{-1}/\sqrt{3}$, where a solid line emits a spatial dimer.
The exponents $n_L(x)$ and $n_T(x)$ in Eq.~(\ref{PARF}) denote the number of T-vertices and L-vertices at spatial position $x$.
In contrast to meson hoppings, spatial baryon hoppings are suppressed in the CT-limit by factors $\gamma^{-1}$. Hence, baryons are static in continuous time. 
Positive (negative) oriented baryons are (dis)favored by a factor $\exp(\pm 3\mu/T)$ over meson lines.
The sign problem has completely vanished!

An important key step towards the CT algorithm is that spatial dimers are distributed uniformly in time as seen in Eq.~(\ref{PARF}). The lengths of dashed or solid intervals (which are related to the number of L- and T-vertices) are, according to a Poisson process, exponentially distributed:
\vspace{2mm}
\begin{equation}
P(\Delta\beta)=\exp(-\lambda \Delta\beta),\qquad \Delta\beta \in [0,\beta=1/aT]
\end{equation}
\vspace{2mm}
with $\lambda$ the ``decay constant'' for spatial dimer emissions. 
Due to the presence of baryons $\lambda$ is space-time dependent: $\lambda=d_M(x,t)/4$, where $d_M(x,t)$ is the number of mesonic neighbors at a given coordinate. 
Non-trivial meson correlations arise from the entropy of the various configurations. 
Likewise, baryonic interactions are due to the modification they induce on the meson bath, and thus also arise from entropy.

The CT algorithm is a Worm-type algorithm, similar to the directed path algorithm introduced for SC-QCD in \cite{Adams1989}. The updating rules are outlined in Fig.~\ref{absorptionemission} and will be explained in detail in a forthcoming publication.

\section{Results on the SC-QCD Phase Diagram}

In SC-QCD at low temperatures, chiral symmetry, i.~e.~the $U_A(1)$ symmetry of the one-flavor staggered action, is spontaneously broken according to $U_L(1) \times U_R(1) \rightarrow U_V(1)$, and becomes restored at some critical temperature $T_c$.


To obtain the SC-QCD phase diagram in the chiral limit $m_q\!=\!0$, we have measured the chiral phase transition temperature as a function of $\mu$, and the nuclear transition with the baryon density. We were able to locate the tricritical point and also find a re-entrance already predicted by mean field analysis, due to the fact that the entropy decreases in the
high-density phase as the lattice becomes saturated with baryons. Our new results eliminate systematic errors affecting previous findings based on mean field approximations \cite{Nishida2004} or Monte Carlo for fixed $\Nt$ \cite{Forcrand2010}.\\

\section{Acknowledgments}

We thank the committee of the QM 2011 poster session to have chosen our poster for the flash talk session.
The computations have been carried out on the Brutus cluster, ETH Z\"urich. This work was supported by the Swiss National Science Foundation under grant 200020-122117.\\

\end{document}